\documentclass[12pt,reqno]{article}
\usepackage[lmargin=2cm, rmargin=2cm]{geometry}
\usepackage{amsfonts}
\usepackage{amssymb}
\usepackage{enumerate}
\usepackage{amsthm}
\usepackage{array}
\usepackage{epsfig}
\usepackage[small,bf]{caption}
\usepackage{appendix}
\usepackage{dsfont}
\usepackage{hyperref}
\usepackage{amsmath}
\usepackage{graphicx}
\usepackage{epstopdf}
\usepackage{blkarray}
\usepackage{subcaption}
\usepackage{cite}

\allowdisplaybreaks \numberwithin{equation}{section}

\begin{document}

\begin{titlepage}
 \thispagestyle{empty}

\begin{flushright}
 \end{flushright}

 \begin{center}

 \vspace{30mm}

     { \LARGE{\bf  {Black hole collapse and democratic models}}}

     \vspace{40pt}

\Large{{\bf Aron Jansen and Javier M. Mag\'an}} \\[8mm]
{\small\slshape
Institute for Theoretical Physics \emph{and} Center for Extreme Matter and Emergent Phenomena, \\
Utrecht University, 3508 TD Utrecht, The Netherlands \\

\vspace{5mm}

{\upshape\ttfamily a.p.jansen@uu.nl, j.martinezmagan@uu.nl}\\[3mm]}

\vspace{8mm}

     \vspace{10pt}

    \vspace{10pt}

\date{\today}

\end{center}

\begin{abstract}
We study the evolution of black hole entropy and temperature in collapse scenarios, finding three generic lessons. First, entropy evolution is extensive. Second, at large times, entropy and temperature ring with twice the frequency of the lowest quasinormal mode. Third, the entropy oscillations saturate black hole area theorems in general relativity. The first two features are characteristic of entanglement dynamics in `democratic' models. Solely based on general relativity and Bekenstein-Hawking entropy formula, our results point to democratic models as microscopic theories of black holes. The third feature can be taken as a prediction for democratic models coming from black hole physics.
\end{abstract}

 \vspace{10pt}
\noindent

\end{titlepage}

\thispagestyle{plain}


\baselineskip 6 mm

\newpage

\section{Introduction}\label{secI}
One of the most influential and paradigmatic results in the context of black hole physics is the discovery of black hole entropy \cite{bekenstein,hawking}:
\begin{equation}\label{entro}
S_{\textrm{BH}}=\frac{A}{4}\;,
\end{equation}
where $A$ is the area of the event horizon in natural units. Although this is believed to be a fundamental relation in quantum gravity, its connection with microscopic physics continues to be obscure. The holographic principle \cite{thooft,hologram} is too generic in this regard, since any theory defined on the event horizon with $\mathcal{O}(1)$ degrees of freedom per Planckian area would saturate~(\ref{entro}), no matter the structure of interactions between them.  It is of obvious interest to ask whether there exist further generic implications of~(\ref{entro}) within potential theories of quantum gravity. In this article, we give evidence that by considering black hole entropy in time-dependent scenarios, one can obtain valuable information about the microscopic structure of the theory. 

Conceptually, the problem we will study is that of entropy production in out of equilibrium (thermalization) processes. Entropy production is a macroscopic thermodynamic quantity directly related to the structure of interactions of the microscopic theory. The reason is that entropy is always generated by some type of coarse graining, defined as a practical inability of measuring a certain type of information (correlators). In out of equilibrium scenarios, deterministic evolution distributes information evenly over all types of correlators \cite{haydeninfo,uscod}, increasing the entropy of the coarse-grained description of the system. Entropy growth is then directly related to the ability of the system to create correlations between the coarse-grained variables and the rest, and these correlations are controlled by the microscopic structure of interactions.

Technically, since black hole entropy is a geometric quantity~(\ref{entro}), the problem is that of studying scenarios with dynamical geometry. A simple possibility is to consider general relativity coupled to a scalar field, and choose an initial state containing a black hole with entropy $S_{\textrm{i}}=A_{\textrm{i}}/4$ and a scalar field profile containing enough energy to backreact on the geometry. As time evolves the scalar field collapses towards the black hole, increasing its entropy to $S_{\textrm{f}}=A_{\textrm{f}}/4$. Given this general context, we will examine concrete examples of spacetimes with different dimensions and scalar fields with different masses, see section~\ref{secII}. We will arrive at three generic lessons, contained in equations~(\ref{extensivelaw}),~(\ref{entropyevo}) and~(\ref{Bmin}).

In section~\ref{secIV} we argue that those lessons contain valuable information about the microscopic interaction structure of black hole dynamics. By reviewing the recent results presented in \cite{usfree}, we will show that the dynamics of black hole entropy perfectly matches the dynamics of entanglement entropy in democratic models. By democratic models we mean the strongest type of non-local models, in which every degree of freedom interacts with every other degree of freedom with couplings of the same size. The importance of such non-local physics in the black hole context was first pointed out in \cite{susskind}.

As we comment below, our results fit well in the specific context of large-$N$ matrix models and the AdS/CFT conjecture \cite{largeN,matrix,adscft}, providing further evidence of the claims presented in \cite{usfree} concerning the entanglement dynamics of large-$N$ matrix models. For the same reasons they nicely connect with the model of black hole dynamics proposed in \cite{sachdev}. Finally, from a different perspective, we also expect our results to contribute to the understanding of the connection between geometry and entanglement (see \cite{takayanagi,vijayhole,myers,sahakian} and references therein).

\section{Entropy production in black holes}\label{secII}
To study the entropy production in black hole collapse scenarios we consider Einstein gravity coupled to a scalar field. The action reads
\begin{equation}
I = \frac{1}{8 \pi} \int d^{d+1} x \sqrt{-g} \left( R + \Lambda - \frac{1}{2} (\partial \phi)^2 - \frac{1}{2} m^2 \phi^2 \right) \, ,
\end{equation}
where $\Lambda = d (d-1)$ is the cosmological constant and the scalar has mass $m$. 
This choice is taken for simplicity and to make direct contact with the AdS/CFT correspondence \cite{adscft}. 

The equations of motion are given by:
\begin{subequations}\label{eq:eom}
\begin{eqnarray}
\label{eq:einst}
0 &=& R_{\mu\nu} - \frac{1}{2} g_{\mu\nu} R  - \frac{1}{2} g_{\mu\nu} \Lambda + \frac{1}{2} g_{\mu\nu} \left( \frac{1}{2} (\partial \phi)^2 + \frac{1}{2} m^2 \phi^2 \right) - \frac{1}{2} \partial_\mu \phi \partial_\nu\phi \, , \\
 \label{eq:scalar} 
0 &=& \partial_\mu( \sqrt{-g} g^{\mu\nu} \partial_\nu \phi ) - \sqrt{-g}\, m^2 \phi \, .
\end{eqnarray}
\end{subequations}
In what follows we take the system to be homogeneous and isotropic at all times. Any such spacetime can be described with the ansatz
\begin{subequations}\label{eq:ansatz}\begin{eqnarray}
ds^2 &=& - f(r,t) dt^2 + 2 dt dr + \Sigma(r,t)^2 d \vec{x}_{d-1}^2 \, , \\
\phi &=& \phi(r,t) \, ,
\end{eqnarray}\end{subequations}
which we choose to simplify the numerical solution later.

The equilibrium solution is a planar AdS-Schwarzschild black hole with temperature $T_f$ and entropy $S_f$,  and with vanishing scalar $\phi = 0$. 
Out of equilibrium, the geometry is asymptotically AdS and the scalar behaves as
\begin{equation} 
\phi(r,t)  = \frac{\phi_0(t)}{r^\Delta}+ ... \, ,
\end{equation}
where $m^2 \equiv \Delta (d - \Delta)$ and dots denote higher order corrections in $1/r$. 

To compute entropy production we assume that~(\ref{entro}) generalises to time-dependent scenarios. Later on, we give some evidence of the validity of this assumption. 
Notice that time-dependence brings a certain ambiguity: we can consider two possibilities for the area appearing in the black hole entropy equation, namely the event horizon area and the apparent horizon area. 
In equilibrium both choices coincide, so it is not clear what is the correct generalisation of~(\ref{entro}) to time-dependent geometries. 
In \cite{ahentropy} it was argued that the apparent horizon area, which is defined as the largest trapped surface, is the correct choice, the main reason being that it can be defined locally in time. 
Below we will study the time evolution of both choices. Although the main features will be shared by both of them, we find more compelling results for apparent horizon area. 
Aside from entropy, we will also associate a temperature to the surface gravity of both horizons.

\subsection{Extensivity of black hole entropy evolution}

From~(\ref{eq:ansatz}), the area of a surface $\mathcal{A}$ at fixed $t$ and $r$ is given by
\begin{equation}
\text{Area}\,(\mathcal{A}) = \int_\mathcal{A} d^{d-1} x \sqrt{-\gamma}\;,
\end{equation} 
with $\sqrt{-\gamma} = \Sigma(r,t)^{d-1}$ the determinant of the spatial part of the metric. Quite strikingly, just from the assumption that the entropy is given by the area of the horizon (apparent or event), it follows directly that given any two disconnected horizon patches $\mathcal{A}$ and $\mathcal{B}$:
\begin{equation}\label{extensivelaw}
S_{\mathcal{A} \cup \mathcal{B}} (t) = S_\mathcal{A} (t) + S_\mathcal{B} (t) \, ,
\end{equation}
black hole entropy is extensive at all times. This implies that the mutual information between different horizon patches vanishes at all times:
\begin{equation}\label{Mutual}
I_{\mathcal{A} \cup \mathcal{B}} (t)\equiv S_\mathcal{A} (t) + S_\mathcal{B} (t) -S_{\mathcal{A} \cup \mathcal{B}} (t)=0\; .
\end{equation}  
Both previous relations are trivial observations from the point of view of gravity, but we will argue in section~\ref{secIV} they are nontrivial from a putative microscopic point of view. Notice that even though we will focus on a homogeneous system, the previous equations (\ref{extensivelaw}) and (\ref{Mutual}) hold generally.

An immediate result of such an extensive law~(\ref{extensivelaw}) is that the characteristic time scale for the stabilisation of the entropy evolution of a certain horizon patch does not depend on the size of the chosen patch. By entropy stabilisation we mean the time by which the entropy evolution enters the plateau regime, where near equilibrium physics hold. We study the physics of the plateau in the next section.

\subsection{Quasinormal ringing of geometric quantities}\label{secIII}
Given the extensive behavior of black hole entropy evolution~(\ref{extensivelaw}), we would like to find the characteristic time scale of near equilibrium relaxation as well as the specific law governing the evolution of geometric quantities at the plateau. 

First we recall that close to equilibrium the scalar field is well described by a sum of quasinormal modes. These are solutions to the linearised equations of motion with ingoing boundary conditions at the event horizon and vanishing Dirichlet conditions at the boundary \cite{quasireview}. These solutions behave as damped harmonic oscillators,
\begin{equation}\label{eq:phi}
\phi(r,t) = A \, e^{- \omega_{\text{I}} t} \left( \,\cos(\omega_{\text{R}} t + \delta) \,\phi_I(r) + \sin(\omega_{\text{R}} t + \delta) \,\phi_R(r)\, \right) \, ,
\end{equation}
as expected for a thermalization process. There is a discrete spectrum of these modes, with higher modes decaying more quickly, so that at late times only the lowest $\omega = \omega_{\text{R}} + i \,\omega_{\text{I}}$ contributes. 
In our setup, the only quasinormal modes that can be excited with the ansatz~(\ref{eq:ansatz}) are those of $\phi$, which we obtain by solving the generalised eigenvalue problem associated to Eq.~(\ref{eq:scalar}) \cite{qnmYaffe,qnmcomputation}. 

Following methods developed in \cite{wilke}, one can study the backreaction  produced by~(\ref{eq:phi}) on the geometry. We first expand the scalar and geometry around equilibrium, $\phi =( \phi_0 =0 )+ \epsilon \delta\phi^{(1)} + \epsilon^2 \delta\phi^{(2)} + ... $ and similar for $f$ and $\Sigma$. 
Plugging this into the equations of motion~(\ref{eq:eom}), at first order in $\epsilon$ we obtain the quasinormal mode equation, there is no backreaction at this order. 
At second order however, $\delta f^{(2)}$ and $\delta \Sigma^{(2)}$ are induced by $\left(\delta \phi^{(1)}\right)^2$.
This implies that at late times, when $\delta\phi^{(1)}$ is the lowest quasinormal mode~(\ref{eq:phi}), the geometry will decay to equilibrium with frequency $2\, \omega_I$, twice as fast as the scalar.

Inspired by this and by the results of \cite{usfree}, we consider the following ansatz for the black hole entropy at the plateau:
\begin{equation}\label{entropyevo}
\delta S(t) \equiv S_{\textrm{f}} - S(t) = A\, e^{-2 \omega_{\text{I}} t} \left(\, \cos(2 \omega_{\text{R}} t + \delta) + B\, \right) \, ,
\end{equation}
where again $\omega$ is the lowest quasinormal mode of the scalar field. 
Here $A$ and $\delta$ parametrise the initial amplitude and phase. The parameter $B$, which we term the \textit{damping shift}, suppresses ($B >1$) or enhances ($B<1$) the oscillations around the decaying exponential.
We consider the same ansatz for the temperature and apply both to the apparent and event horizon.

Notice that entropy evolution is special in one regard. Due to area theorems in general relativity, we have:
\begin{equation}\label{eq:increase}
S'(t) \geq 0 \, ,
\end{equation}
implying the following constraint on the damping shift:
\begin{equation}\label{Bmin}
B \geq B_{\text{min}} \equiv \frac{\sqrt{\omega_\text{I}^2 + \omega_\text{R}^2}}{\omega_\text{I}}\, .
\end{equation}

To determine whether the ansatz~(\ref{entropyevo}) provides a good description of the time evolution, and to find the parameters in the ansatz, we must solve the full non-linear equations of motion~(\ref{eq:eom}).
We do this numerically following the methods of \cite{chesleryaffe}.
This approach requires a choice of an initial profile for the scalar, which we take to be  $\phi(t=0,r) = T / r^\Delta$, but we stress that our results do not depend on this choice.

\begin{figure}[h]
\begin{center}
\includegraphics[scale=.45]{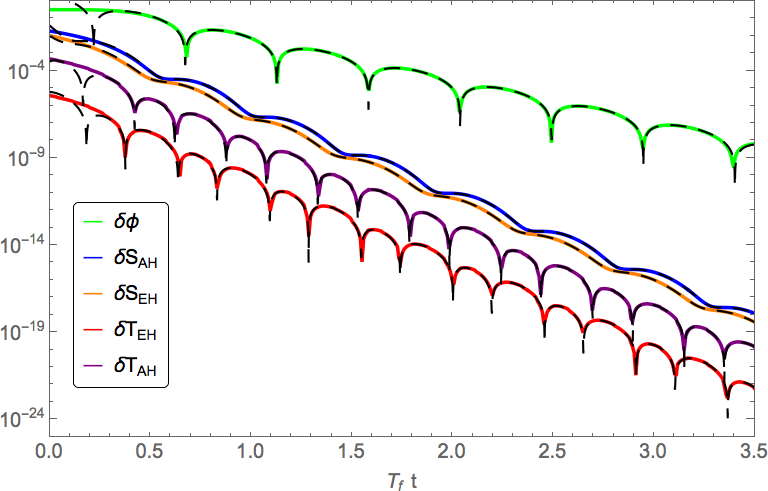}
\end{center}
\caption{Logarithmic plot of the near equilibrium evolution of the geometry in response to a small scalar perturbation, for $d=4$ and $\Delta = 3$.
Shown, from top to bottom, are the scalar field (Green), the areas of the apparent and event horizon (Blue, Orange), and the temperatures of the event and apparent horizon (Red, Purple). 
The $\delta$ always refers to the final value minus the current value, and the last two quantities are shifted down by $10^{-1}$ and $10^{-3}$ respectively for display purposes.
Through each of these quantities a fit of the form of Eq.~(\ref{entropyevo}) is plotted as a dashed line.}
\label{logplot}
\end{figure}

The results are shown in Fig~\ref{logplot}, where we show the evolution of the entropy and temperature derived from both event and apparent horizons for $d=4$ and $\Delta = 3$.
Dashed lines are fits of the form~(\ref{entropyevo}) to each of these quantities, showing that this ansatz accurately describes the evolution. Actually we find that the full geometry, $\delta f(r,t)$ and $\delta\Sigma(r,t)$, are described by the ansatz of Eq.~(\ref{entropyevo}), where now the parameters $A$, $B$, $\delta$ depend on $r$.

By studying different initial profiles we see that the parameters $A$ and $\delta$ depend on the initial conditions, but the damping shift $B$ does not.
We repeat the same process for a range of different dimensions and masses, obtaining in each case a picture which is qualitatively the same as Fig~\ref{logplot}. In Table~\ref{paramvalues} the damping shift $B$ is shown for each quantity in all the studied cases.

\begin{table}
\centering
\begin{tabular}{c ||| c|c||c|c|c|c|}
  $(d,\Delta ) $ & $ \omega / \pi T $ & $ B_\text{min} $ & $ B_{\textrm{SAH}} $ & $ B_{\textrm{SEH}} $ & $ B_{\textrm{TAH}} $ & $ B_{\textrm{TEH}} $ \\ \hline\hline
 $(3,2) $ & $ 1.6372+2.0444 i $ & $ 1.281134 $ & $ 1.281134 $ & $ 1.4547 $ & $ 0.464 $ & $ 0.464 $ \\ \hline
 $(3,3) $ & $ 2.4659+3.5518 i $ & $ 1.217376 $ & $ 1.217377 $ & $ 1.3845 $ & $ 0.532 $ & $ 0.532 $ \\ \hline
 $(4,3) $ & $ 2.1988+1.7595 i $ & $ 1.600513 $ & $ 1.60051 $ & $ 2.0465 $ & $ 0.194 $ & $ 0.236 $ \\ \hline
 $(4,4) $ & $ 3.1195+2.7467 i $ & $ 1.513227 $ & $ 1.51323 $ & $ 1.97 $ & $ 0.196 $ & $ 0.285 $ \\ \hline
 $(5,3) $ & $ 1.6165+0.8419 i $ & $ 2.164801 $ & $ 2.1648 $ & $ 2.880 $ & $ 0.0314 $ & $ 0.640 $ \\ \hline
 $(5,4) $ & $ 2.4574+1.4855 i $ & $ 1.933039 $ & $ 1.9330 $ & $ 2.718 $ & $ 0.0108 $ & $ 0.905 $ \\ \hline
 $(5,5) $ & $ 3.3087+2.1547 i $ & $ 1.832485 $ & $ 1.8325 $ & $ 2.7 $ & $ 0.00994 $ & $ 1.10 $ \\ \hline
\end{tabular}
\caption{Fit results of the damping shift $B$  of Eq.~(\ref{entropyevo}) for the entropy and temperature as defined by the apparent horizon and event horizon, for a range of different dimensions $d$ and scaling dimensions $\Delta$.}\label{paramvalues}
\end{table}

Remarkably, the evolution of the apparent horizon area saturates the bound of Eq.~(\ref{eq:increase}) at each period of oscillation, as can be seen qualitatively from the blue line in Fig~\ref{logplot} and quantitatively for each case by the equality $B_{\textrm{SAH}} = B_{\text{min}}$ from Table~\ref{paramvalues}. The oscillations around the decaying exponential are maximal and saturate the area theorem in general relativity.

On the other hand, the oscillations of the temperature are generally enhanced with respect to the scalar field, i.e. $B_{\textrm{T}} < 1$. Here again, the temperature of the apparent horizon oscillates more strongly than that of the event horizon, with the difference becoming much more pronounced for larger dimensions.

Finally, we would like to note that by adding a constant source for the scalar, backreaction occurs at first order. This changes the factor $2\,\omega_{\text{I}}$ in~(\ref{entropyevo}) to $\omega_{\text{I}}$, but we check that the apparent horizon still saturates the bound~(\ref{eq:increase}), the damping shift being given by~(\ref{Bmin}).

Summarizing, we arrived at three generic lessons. First, the evolution of black hole entropy is extensive~(\ref{extensivelaw}). This implies that the evolution of any subsystem enters the near equilibrium regime in a time scale which is independent of the system size, just dependent on the lowest quasinormal mode. Second, the evolution at the plateau is accurately described by the ansatz~(\ref{entropyevo}). Third, when measured by the apparent horizon, black hole entropy evolution saturates area theorems, allowing us to extract an analytical formula for the damping shift $B$ appearing in~(\ref{entropyevo}).

\section{Entropy evolution in democratic models}\label{secIV}

In the previous section we studied the evolution of black hole entropy in general relativity, arriving at three generic lessons. As for the static Bekenstein-Hawking formula~(\ref{entro}), we expect these lessons to contain information about the microscopic structure of quantum gravity.

The first question we want to ask is: what entropy is the horizon area computing in time-dependent scenarios? A natural proposal is the following:
\begin{equation}\label{Delta}
\Delta S(t) \equiv S(t) - S_{i} =  \Delta S_{\textrm{BH}_{\textrm{in}}}(t)+\Delta S_{\varphi}(t)\, ,
\end{equation}
where $\Delta S_{\textrm{BH}_{\textrm{in}}}(t)$ accounts for the change in entropy of the degrees of freedom conforming the intial black hole and $\Delta S_{\varphi}(t)$ accounts for the change in entropy of the microscopic degrees of freedom supporting the scalar field. Notice that the previous relation must hold at stationarity, and it is certainly natural to assume it in the time-dependent scenario. 
Given that entropy in unitary quantum mechanical scenarios emerges as entanglement entropy \cite{lloyd,page,usrandom,eisert}, the variation $\Delta S_{\varphi}(t)$ should account for the entanglement entropy between the scalar field and the degrees of freedom supporting the initial black hole.
Writing $\delta S_{\textrm{BH}_{\textrm{in}}}(t) \equiv\Delta S_{\textrm{BH}_{\textrm{in}}}(t_f)- \Delta S_{\textrm{BH}_{\textrm{in}}}(t)$ and $\delta S_{\varphi}(t)\equiv\Delta S_{\varphi}(t_f)-\Delta S_{\varphi}(t)$,
relation~(\ref{Delta}) can be written as:
\begin{equation}\label{delta}
\delta S(t) = S_{\textrm{f}} - S(t) =  \delta S_{\textrm{BH}_{\textrm{in}}}(t)+\delta S_{\varphi}(t)\, .
\end{equation}
Since both terms in the right hand side are positive, we conclude that $\delta S(t)$ is an upper bound for both terms separately. Therefore, black hole entropy production should bound the approach of the entanglement entropy between the scalar field and the black hole, as quantified by $\delta S(t)>\delta S_{\varphi}(t)$. 
But notice that the deviation from stationarity of black hole entropy is controlled by the time scales associated to the scalar field, and not by any internal properties. This observation suggests that the classical gravitational description misses $\delta S_{\textrm{BH}_{\textrm{in}}}(t)$, providing:
\begin{equation}\label{eq}
\delta S(t) = A \, e^{-2 \omega_{\text{I}} t} \left(\, \cos(2 \omega_{\text{R}} t + \delta) + B \,\right) = \delta S_{\varphi}(t)\;.
\end{equation}
It seems that to get $\delta S_{\textrm{BH}_{\textrm{in}}}(t)$ we have to resort to the microscopic theory, an interesting observation worth exploring further.

To avoid possible confusion, we want to remark that our results are not in contradiction with the known evolution of spatial entanglement entropy in AdS/CFT, see \cite{calabresecardy,esperanza,vijay}. Here we are not considering `spatial' entanglement entropy of a putative dual field theory. The claim is that we are bounding the entanglement between the initially out of equilibrium fields and the black hole.

The second question we would like to ask is: what are the class of models satisfying such entropy dynamics? A specific class of many body quantum theories displaying the scaling behavior~(\ref{extensivelaw}) and quasinormal decay~(\ref{entropyevo}) for the dynamics of their entanglement entropies have recently been found in \cite{usfree}. This is the class of `democratic' theories, in which every oscillator interacts with every other oscillator through terms in the Hamiltonian. For these theories the entanglement entropy evolution of a set $A$ of oscillators, labelled by $i=1,\cdots, M$ was found to satisfy:
\begin{equation}\label{extensivity}
S_{A}(t)=\sum\limits_{i=1}^{M}S_{i}(t)\;,
\end{equation}
where
\begin{equation}
S_{i}(t)=(n_{i}(t)+1)\log (n_{i}(t)+1)-n_{i}(t)\log n_{i}(t)\;,
\end{equation}
and $n_{i}=\langle \,a_{i}^{\dagger}a_{i}\rangle$ is the average occupation number of the i-th oscillator. The previous result is valid up to subleading corrections in the thermodynamic limit. It mirrors~(\ref{extensivelaw}) and it again implies that mutual information vanishes in the thermodynamic limit at all times as in~(\ref{Mutual}). Besides, on general grounds the behavior of the occupation number is that of the square of the corresponding oscillator, since we need to construct a product of creation and annihilation operators. Therefore, it was argued in \cite{usfree} that $n_{i}$ would ring at the plateau with twice the quasinormal frequency of the lowest quasinormal mode of subset $A$:
\begin{equation}\label{decay}
n_{i}(t)\simeq n_{i}^{\beta}+\, A\,e^{-2\,\omega_{\text{I}}\,t}\,(\,\cos (2\,\omega_{\text{R}}\,t\,+\delta)\, +1\,)\;,
\end{equation}
implying an analogous behavior for the entropy evolution. The behavior predicted in \cite{usfree} for democratic systems perfectly matches the black hole results~(\ref{entropyevo}) and~(\ref{eq}) at large times.

From a different perspective, these results fit remarkably well in the context of large-$N$ matrix models and AdS/CFT \cite{largeN,matrix,adscft}. 
As first remarked in \cite{susskind}, the color dynamics of large-$N$ matrix models is an example of a democratic system. 
In this context, black hole entropy production would bound the entanglement between the out of equilibrium operators supporting the scalar field and the rest of operators as time evolves, i.e the entanglement between different subsectors of the matrix model. 
Our results give further evidence for the claims presented in \cite{usfree}, where it was argued that relation~(\ref{extensivity}) is equivalent to the famous large-$N$ factorisation \cite{largeN}, and where~(\ref{entropyevo}) was found to describe this special type of entanglement dynamics.

\section{Conclusions}

In the first part of this article we studied black hole entropy production. Since black hole entropy is a geometric quantity~(\ref{entro}) we considered scenarios with dynamical geometry. We found three general lessons:
\begin{itemize}
\item Black hole entropy evolution is extensive~(\ref{extensivelaw}). The characteristic time scale for entering the near equilibrium regime is independent of the system size. 
\item At the entropy plateau, black hole entropy and temperature ring with twice the frequency of the lowest quasinormal mode~(\ref{entropyevo}).
\item For the apparent horizon area the damping shift $B_{\textrm{SAH}}$ is the maximal one compatible with area theorems in general relativity~(\ref{Bmin}).
\end{itemize}
These universal macroscopic lessons are interesting in their own right and are part of the main results of the article.

In the second part of the article, we provided a coherent microscopic interpretation of the first two macroscopic results. 
Firstly, we argued on general grounds that the deviation from stationarity of black hole entropy should bound the deviation from stationarity of the entanglement between the out of equilibrium modes associated to the classical scalar field and the initial black hole degrees of freedom. 
Secondly, we presented a class of systems with such highly non-trivial entanglement dynamics. This is the class of democratic systems, see section~\ref{secIV} and \cite{usfree}. Given this connection, the fact that the damping shift $B_{\textrm{SAH}}$ saturates area theorems in general relativity is an unexpected prediction from black hole physics to the physics of democratic systems.

We end this article with some remarks. 
The first is that these results have been obtained from a purely macroscopic perspective. We have not invoked at any time any specific microscopic theory of quantum gravity, such as string theory \cite{stringbooks}. 
It is tempting to conclude that any putative theory of quantum gravity should show such democratic behavior, whether at a microscopic or at an emergent level. 
On the other hand, string theory complies with such requirements, as first acknowledged in \cite{susskind}. 
Indeed, within matrix models and AdS/CFT \cite{largeN,matrix,adscft} our results seem to fit remarkably well. 
In this context, the deviation from stationarity of black hole entropy bounds the entanglement dynamics between different subsectors of the large-$N$ matrix model. 
In the field theory, this type of entanglement dynamics were studied in \cite{usfree}, where the behavior~(\ref{entropyevo}) was found. 
This unexpected agreement between both computations provides further evidence of the claims made in \cite{usfree}, concerning the fast dynamics of entanglement growth in large-$N$ matrix models, and deserves further consideration. 
For the same reasons, our results might also be interesting for the microscopic approach to black hole dynamics presented in \cite{sachdev}.

We also want to remark that the observed damping shift $B$ deserves further consideration. It would be interesting to obtain analytical control over it in the case of other geometric quantities like the temperature and also obtain a microscopic understanding of the mechanism that gives rise to the saturation~(\ref{Bmin}). It would be very interesting to see if this saturation is characteristic of democratic models.

Another interesting application of our results concerns the dynamics of spatial entanglement entropy in AdS/CFT \cite{esperanza,vijay}. Given that at sufficiently large times the minimal area surface has a contribution coming from the area of the event horizon, we expect to see the quasinormal oscillations in such a quantity as well. These oscillations were missed in those references due to the specific nature of the Vaidya metric, and we expect to see them in more realistic collapse scenarios.

Lastly, we hope that the relation between horizon area, quasinormal ringing and entanglement dynamics uncovered in this article will contribute to a better understanding of the puzzle of deriving geometry from the properties of entanglement (see \cite{takayanagi,vijayhole,myers,sahakian} and references therein). Potentially, it might also be of interest to approaches of emergent general relativity based on entropy densities of holographic screens \cite{jacobson,verlinde}.

\section*{Acknowledgements}
We thank Marcos Crichigno, Umut G\"ursoy, Esperanza Lopez, Wilke van der Schee, Phil Szepietowski and Stefan Vandoren for interesting discussions. This work was supported by the Delta-Institute for Theoretical Physics (D-ITP) that is funded by the Dutch Ministry of Education, Culture and Science (OCW).


\end{document}